# Second harmonic generation in a centrosymmetric gas medium with spatiotemporally focused intense femtosecond laser pulses

Guihua Li,[1,3] Jielei Ni,[1] Hongqiang Xie,[1,3] Bin Zeng,[1] Jinping Yao,[1] Wei Chu,[1] Haisu Zhang,[1,3] Chenrui Jing,[1,3] Fei He,[1] Huailiang Xu,[2,†] Ya Cheng,[1,#] and Zhizhan Xu[1,*]

[1]State Key Laboratory of High Field Laser Physics, Shanghai Institute of Optics and Fine Mechanics, Chinese Academy of Sciences, Shanghai 201800, China

[2]State Key Laboratory on Integrated Optoelectronics, College of Electronic Science and Engineering, Jilin University, Changchun 130012, China

[3]University of Chinese Academy of Sciences, Beijing 100049, China

[†] huailiang@jlu.edu.cn
[#] ya.cheng@siom.ac.cn
[*] zzxu@mail.shcnc.ac.cn

**Abstract.** We demonstrate unexpectedly strong second harmonic generation (SHG) in Argon gas by use of spatiotemporally focused (SF) femtosecond laser pulses. The resulting SHG by the SF scheme at a 75 cm distance shows a significantly enhanced efficiency than that achieved with conventional focusing scheme, which offers a new promising possibility for standoff applications. Our theoretical calculations reasonably reproduce the experimental observations, which indicate that the observed SHG mainly originates from the gradient of nonuniform plasma dynamically controlled by the SF laser field.

PACS numbers: 42.65.Ky, 42.65.Re

# 1. Introduction

When an intense femtosecond laser pulse is focused into gas phase atoms or molecules, harmonics can be easily generated. In the perturbative regime, only the low-order ($q$-th) harmonics can be observed because the conversion efficiency of harmonics is proportional to $q$-th power of incident laser field [1,2]. In the nonperturbative regime, high-order harmonic generation (HHG), whose spectrum is featured with a spectacular plateau with a sharp cutoff, has attracted tremendous attention since its first observation by McPherson et al. in 1987 [3,4]. For the reason of symmetry, typically only odd-numbered harmonics can be produced with noticeable efficiency in centrosymmetric media such as monatomic or unaligned molecular gases.

Surprisingly, though it appears to be counterintuitive, second harmonic generation (SHG) has indeed been observed in air with intense femtosecond laser pulses [5,6]. With some careful examinations, the mechanism behind such phenomenon is attributed to the gradient of photoionization-induced plasma caused by the pondermotive force [5]. However, it was noted that high conversion efficiency of SHG can only be achieved in tightly focused laser schemes [5,7]; whereas in a loose focusing scheme [6], because of the limited laser intensity and plasma density resulted from the defocusing effect in the laser-induced plasma (intensity clamping) [8,9], the evaluated conversion efficiency of SHG is as low as $10^{-7}-10^{-6}$. In addition, the detection of second harmonic signals at a remote distance becomes even more difficult because of the strong background of white light generation in a femtosecond filament [10]. Such significant drawbacks hamper the use of this technique as an efficient means for plasma diagnostics [11].

In this work, we show that the above-mentioned difficulties can be overcome with spatiotemporally focused (SF) femtosecond laser pulses. The trick of the SF technique is to separate the spectral components of a femtosecond pulse in space before the pulse enters the objective lens. A temporal focusing occurs because the spatial overlapping of different frequency components only happens around the focal point, leading to the shortest pulse duration and consequently, the highest peak intensity [12,13]. Previously, we have shown that with the SF scheme, extremely short filament can be produced in air at a far distance with a significant enhancement of the peak intensity despite of the strong plasma defocusing effect [14]. Here, we find that with the SF femtosecond laser beam, it is possible to control the plasma dynamics at femtosecond time scale, which leads to more efficient SHG as compared with conventional focusing (CF) excitation scheme using a transform-limited Gaussian beam. The underlying mechanism for such remotely strong SHG by the SF scheme is ascribed to the gradient of nonuniform plasma induced by the intense SF laser field, which however shows a completely different plasma dynamic behavior from that induced by the CF laser field.

## 2. Experimental

As illustrated in Fig. 1, our experiment was carried out with a commercial Ti:sapphire laser system (Legend Elite Duo, Coherent, Inc.), which delivers ~6-mJ pulses with a center wavelength at ~800 nm and a spectral bandwidth of ~30 nm at a repetition rate of 1 kHz. The amplified but uncompressed laser pulses emitted from the laser system were first spatially dispersed along the horizontal direction (i.e., X axis as shown in Fig. 1) by a pair of 1500 lines/mm gratings parallel to each other (blazing at ~50°). The distance between the two gratings was adjusted to be ~80 cm to compensate for the temporal dispersion of different spectral components of the uncompressed laser pulses. After the gratings, the laser beam became elliptical due to the spatial dispersion, i.e., its diameters were measured to be ~40 mm ($1/e^2$) and ~9 mm ($1/e^2$) along X-axis (i.e., spatial chirp direction) and Y-axis directions, respectively. A half-wave plate was inserted before the gratings to adjust the polarization direction of the fundamental beam. The laser pulse of ~1.0 mJ was then focused by a 75-cm-focal-length lens into a chamber filled with Ar gas at a pressure of 1 bar. The generated second harmonic radiation was recorded by a spectrometer (Shamrock 303i, Andor Corp.) after its beam size being approximately reduced by half with a combination of a convex lens and a concave lens [Lens 2 ($f$= 25 cm) and Lens 3 ($f$= -12 cm), respectively]. The spatial profile of the second harmonic beam in the focal plane was imaged by a convex lens (Lens 4: $f$= 20 cm) on an area CCD camera (Wincamd-ucd23). Three dichroic mirrors (DM1, DM2, DM3) with high reflectivity at 400 nm and high transmission at 800 nm were used to separate the fundamental beam from the second harmonic radiation. A Glan-Taylor prism before the spectrometer was used to measure the polarization of SHG. For comparison, SHG experiment with a CF scheme was also carried out by removing the grating pair while keeping all the other conditions unchanged. In this case, the transform-limited incident pulse was measured to have a pulse duration of ~40 fs with a nearly circular beam profile of a diameter of ~9 mm ($1/e^2$).

## 3. Experimental results

Figure 2 shows various properties of the second harmonic beam from Ar gas by use of SF femtosecond laser pulses. When the polarization direction of fundamental laser was set along Y axis (SF-PY), the focus of the SHG beam, which is obtained from the CCD in Fig. 1, shows two lobes separated from each other along the Y-axis, which is coincident with previous investigations by using tightly focused laser scheme [5,7]. Nevertheless, when the polarization direction of fundamental laser was set along X axis (SF-PX), i.e., the spatial dispersion direction of the incident beam before the focal lens, the second harmonic beam profile becomes a Gaussian-like distribution without any splitting, as shown in Fig. 2(b).

The second harmonic spectra of Fig. 2(a) and 2(b) are also illustrated in Fig. 2(c) by the blue-solid and red-dashed curves, respectively. For comparison, the spectrum generated by the CF scheme with the fundamental laser polarization along X axis (CF-PX) is also shown in Fig. 2(c) by the green-dot line. We found that for the experimental conditions in Fig. 2(a), the measured spectra are centered at ~400 nm, which confirms that the second harmonic of the fundamental laser pulse was indeed generated. Note that the narrow peaks appearing in Fig. 2(c) are fluorescence emissions from Ar and very small amount of $N_2$ in the chamber due to the leakage

[15]. Furthermore, it is found that by only changing the fundamental laser polarization, as shown in Fig. 2(b), the efficiency of SHG is significantly enhanced by about 5 times than that obtained with the pump laser polarized along Y-axis. Meanwhile, it can be seen from Fig. 2(c) that the SHG with the CF scheme (green dot line) is too weak to be distinguished from the tail of a strong supercontinuum white light, which has been generated in the filamentation process due to self-steepening and self-phase modulation [16]. Since the white light develops progressively along the plasma filament, the much shorter filament length of ~1.5 mm produced with the SF scheme results in the dramatically reduced (almost unobservable) white light spectra as shown by the red and blue curves in Fig. 2(c). As a result, the SF laser pulses can generate second harmonic efficiently with an extremely high signal-to-noise ratio, especially when the pump laser is linearly polarized along the direction of the spatial chirp (i.e., along X-axis direction).

The polarization property of the second harmonic radiation provides direct information to clarify the origin of the SHG. We carefully checked the polarization property of the second harmonic beam obtained with the SF scheme, and found that for the SF-PX case, the second harmonic is linearly polarized in the direction parallel to that of the fundamental laser, as shown in Fig. 2(d). The same behavior has been observed for the SF-PY case (not shown) as well. Thus, we can exclude the contribution from the third-order nonlinear mixing via atomic nonlinear susceptibility ($\chi^3$) between the laser field and the ionization-induced electric field, since such nonlinear mixing would result in a radial polarization of second harmonic [17-19]. As we show below, the strong second harmonic signal still originates from the gradient of photoionization-induced plasma, whereas the plasma behavior in the SF laser fields is completely different from that in the CF laser fields.

## 4. Theoretical analysis

Our calculations of the SHG are performed based on the approach of free-electron second-order susceptibility [17,20],

$$\vec{P}(2\omega) = \frac{n_e e^3}{8m^2\omega^4}\nabla E^2 + \frac{e\vec{E}(\nabla \cdot \vec{E})}{8\pi m\omega^2}, \qquad (1)$$

where $n_e$ represents the density of plasma ionized by the driver laser field with the frequency $\omega$. The first term comes from the current of electrons driven by the ponderomotive force, and the second one arises from the gradient of nonuniform plasma. Since in Eq. (1), the first term would result in second harmonic with an isotropic polarization distribution, whereas the second term lead to second harmonic with the polarization direction parallel to that the fundamental laser [5,7,17,21], the second term should dominate the SHG in the SF scheme according to the experimental observation shown in Fig. 2(d). Therefore, in the following, we will only focus on the second term.

First, we calculate the electron density variation $\delta n$, which is directly related to the SHG process. In our calculations, we assume that the initial electron density produced by photoionization is uniform ($n_0$) [5,7] and the ponderomotive force is responsible for the inhomogeneity of the plasma density by taking the electron dynamics in account as below

$$\frac{d\vec{v}}{dt} = k\nabla I(\mathrm{r},\mathrm{t}), \qquad (2)$$

where $k$ is a constant. Then we can obtain $\delta n$ based on the continuity equation [20],

$$\frac{d\,\delta n}{dt} = -n_0 \nabla \cdot \vec{v}. \qquad (3)$$

Since the light field in the femtosecond laser pulses changes with time, the calculated $\delta n$ is also a function of time. Here, to provide a clear physical picture, we only show the calculated $\delta n$ in the focal plane at the end of the driving pulse, i.e., at $t$=100fs, for either the SF [Fig. 3(a)] or the CF [Fig. 3(b)] laser fields. It is noteworthy that the distribution of $\delta n$ is independent of the polarization direction of the pump laser pulse. It is found that changing the parameters such as the laser peak intensity and focal spot size will only slightly change the calculated results, whereas all the major features will still remain. Since obtaining exact values of the parameters such as the peak intensity and the focal spot size by experimental measurements is difficult, we tuned these parameters in our numerical calculation until a good agreement between the experimental and theoretical results has been eventually achieved. It can be clearly seen that in Fig. 3(b), the calculated $\delta n$ is centrosymmetric and the minimum is at the central part, which can be easily understood by the ponderomotive force driven by the CF Gaussian laser field. In the central area of the CF laser beam, the laser intensity is nearly uniformly distributed, thus there is almost no gradient in the plasma density. Surprisingly, with the SF scheme, electrons are strongly pushed to one side along the X axis by the pondermotive force as evidenced in Fig. 3(a), which stems from the unique laser field described by [22],

$$E(z=f,t) \propto \exp[-\Omega^2(t+\alpha xk/f)^2/4] \qquad (4)$$

where $\alpha$ represents the spatial chirp and $\sqrt{2}\Omega$ is the spectral bandwidth at $1/e^2$. Equation (4) shows that the laser field with SF schemes can give rise to a significant pulse-front tilt along the X-axis direction (i.e., shortest pulses arrive at different times with a time shift $t_0 = \alpha kx/f$ along X direction). The influence of the ponderomotive force on the plasma dynamics is greatly enhanced by the pulse-front tilt, forcing the major part of electrons to be distributed at the edge of the spatiotemporally focused spot as shown in Fig. 3(a).

With the calculated electron density variation $\delta n$, the source of SHG, i.e., the current density $\vec{J}_{2\omega}$ of all the quivering electrons can be expressed as $\vec{J}_{2\omega} = \delta n_q e \vec{v}_{quiver}$, where $\vec{v}_{quiver} = e\vec{E}/(im\omega)$ is the velocity of the quivering electrons and $\delta n_q = \nabla \delta n \cdot \vec{v}_{quiver}$ is the charge perturbation by the quivering electrons. The SHG can be calculated using the following expression,

$$\vec{P}(2\omega) = \frac{\vec{J}_{2\omega}}{-2i\omega} = \frac{e^3}{2im^2\omega^3}(\nabla \delta n \cdot \vec{E})\vec{E}. \qquad (5)$$

It can be seen from Eq. (5) that $\vec{P}(2\omega) \propto \nabla \delta n \cdot \vec{e}$, where $\vec{e}$ is the unit vector of $\vec{E}$. This equation again manifests that SHG is proportional to the gradient of plasma density along the polarized direction.

Based on Eq. (5), we theoretically calculate the SHG in the focal plane using our experimental parameters, and the results are shown in Fig. 4. It is noteworthy that the SHG results are averaged temporally over the whole pulse duration. It can be seen in Fig. 4 that our simulations successfully reproduce all the major features of the experimentally measured second harmonic patterns [see, Figs. 2(a) and 2(b)], that is, the SF laser polarized along Y-axis produces a double-lobe second harmonic beam separated in Y direction [Fig. 4(a)] and SF laser polarized along X-axis generates a main lobe of Gaussian-like second harmonic beam accompanied by a much weaker satellite second harmonic beam. The satellite second harmonic signal is unobservable in our measurement due to the limited sensitivity of the CCD detector [Fig. 2 (b)]. On the other hand, it can be observed in Fig. 4(c) that with the CF scheme, a calculated double-lobe pattern featured with two lobes distributed along the polarization direction, is in good agreement with previous measurements [5,7]. For such a situation, owing to the rotationally symmetric $\delta n$, $\nabla \delta n \cdot \vec{e}$ approaches zero on the axis perpendicular to the polarized direction, giving rise to the splitting along the polarized direction.

With our calculation, we also compared the SHG efficiency obtained with the two pump laser polarization directions in the SF scheme, as illustrated in Fig. 4(d). It can be observed that the SHG efficiency with SF laser polarized along X axis is more than 2 times higher than that with the SF laser polarized along Y-axis. This qualitatively agrees with our experimental observation as shown in Fig. 2(c). The quantitative difference should result from the fact that in our calculation, only the SHG at the focal plane was considered, i.e., all the propagation effects have been ignored. To include these effects, a complete modeling of the laser propagation in the plasma should be carried out, which will be considered in our future work.

## 5. Conclusion

In conclusion, we have investigated both experimentally and theoretically the SHG in centrosymmetric Ar gas by use of SF linear-polarized femtosecond laser pulses. We observed a splitting in the SHG beam with the pump laser polarization direction perpendicular to that of the spatial dispersion, and a much stronger Gaussian-like SHG beam with the pump laser linearly polarized along the dispersion direction. By examining the properties of second harmonic radiation including its spatial profile, spectrum and polarization, we found that all the experimental results can be well explained based on the electron density variation driven by the pondermotive effect. Owing to the peculiar density variation with Y-symmetry and X-asymmetry profiles, the gradient of inhomogeneous plasma density mainly contributes to X (spatial dispersion) direction, which causes higher efficiency of SHG driven by X-polarized SF laser than that of Y-polarized SF laser. Therefore, the gradient of nonuniform plasma dynamically controlled by a SF laser beam is regarded to be responsible for the SHG in centrosymmetric madia. More specifically, spatial chirp-induced pulse-front tilt plays a crucial role in such process. Our results show that the SF method provides a promising way for controlling plasma dynamics in a light focal spot at the femtosecond time scale, which may find more useful applications other than the SHG in centrosymmetric media.

**Acknowledgments**

Guihua Li and Jielei Ni contribute equally to this work. This work is supported by the National Basic Research Program of China (Grants No. 2011CB808102 and No. 2014CB921300), and National Natural Science Foundation of China (Grants No. 11127901, No. 11134010, No. 61221064, No. 61275205, No. 61235003 and No. 11204332), the Program of Shanghai Subject Chief Scientist (11XD1405500), the Open Fund of the State Key Laboratory of High Field Laser Physics (SIOM) and the Fundamental Research Funds of Jilin University.

**Figure 1:**

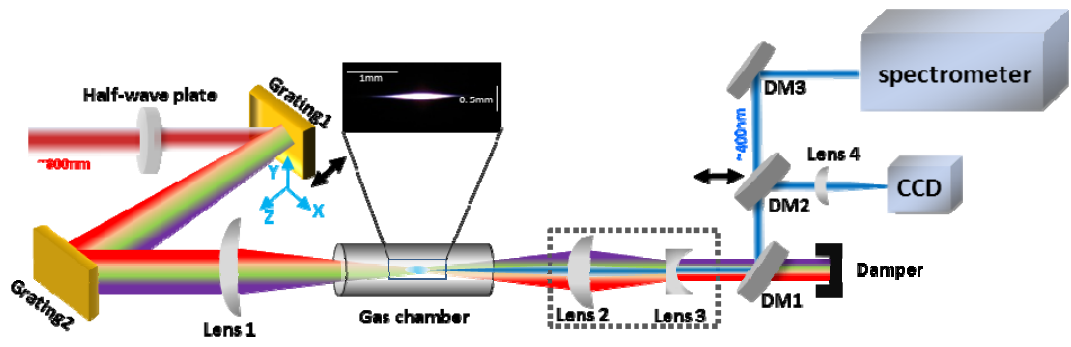

Fig. 1. (Color online) Schematic of the experimental setup

**Figure 2**

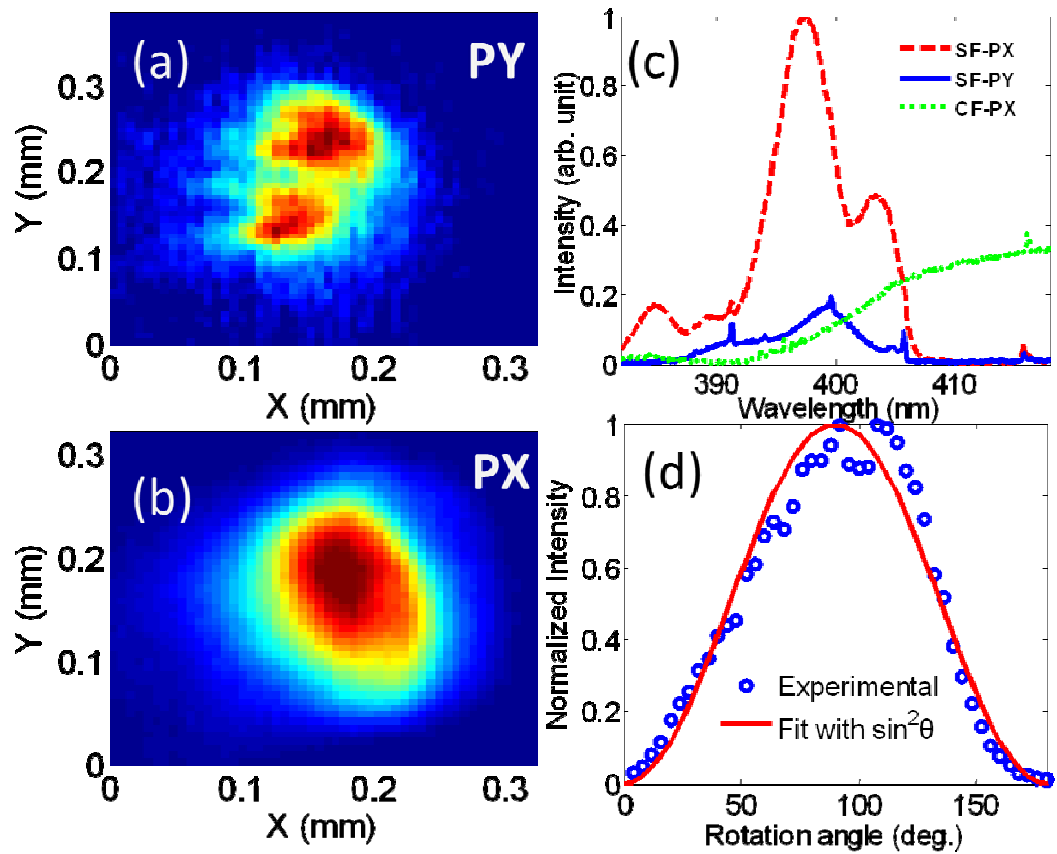

Fig. 2. (Color online) The spatial patterns of SHG by the SF scheme with the fundamental laser linearly polarized (a) along Y axis (SF-PY) and (b) along X axis (SF-PX); (c) The second harmonic spectra obtained in SF-PY (blue-solid line), SF-PX (red-dashed line) and the CF (green-dot line) schemes; (d) The measured polarization property of SH for the SF-PX case (blue circle) and the fit (red line) with $\sin^2\theta$.

**Figure 3:**

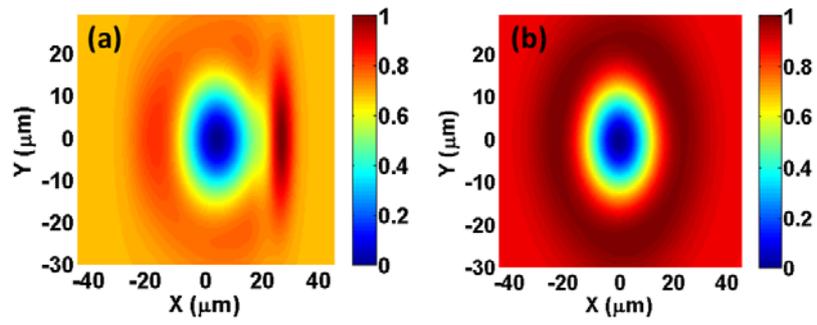

Fig. 3. (Color online) Calculated normalized electron density variation driven by linearly polarized laser with (a) SF scheme and (b) CF scheme.

**Figure 4:**

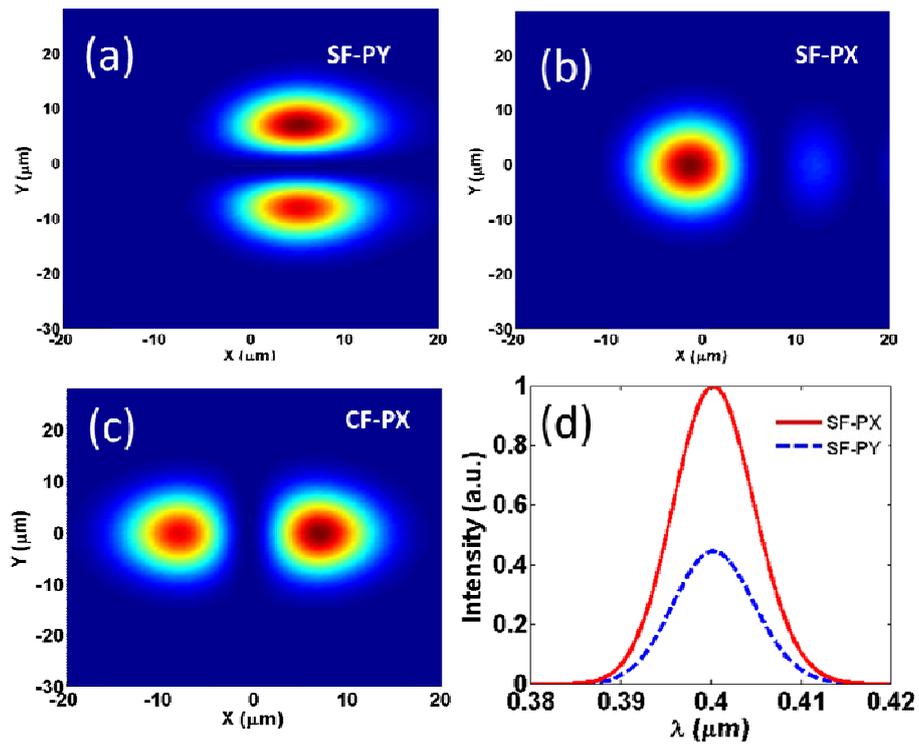

Fig. 4. (Color online) The spatial SH patterns simulated (a) in SF-PY, (b) SF-PX and (c) CF-PX; (d) The calculated second harmonic spectra with SF by the fundamental light polarized along X axis (red-solid line) and along Y axis (blue-dashed line).